\documentclass[prl,twocolumn,showpacs,amsmath,letter]{revtex4}
\usepackage{graphicx}

\newcommand{\Journal}[4]{#1 \textbf{#2}, #3 (#4)}

\begin{document}

\title{Effects of Antiferromagnetic Spin Rotation on Anisotropy of Ferromagnetic/Antiferromagnetic Bilayers}
\author{S. Urazhdin}
\author{C.L. Chien}
\affiliation{Department of Physics and Astronomy, Johns Hopkins
University, Baltimore, MD, 21218}


\begin{abstract}

In epitaxial (111) oriented Ni$_{80}$Fe$_{20}$/Fe$_{50}$Mn$_{50}$ bilayers, we separate two distinct behaviors: unidirectional anisotropy (exchange bias) in thick Fe$_{50}$Mn$_{50}$, and enhanced coercivity in thin Fe$_{50}$Mn$_{50}$. By measuring the magnetization response to a rotating magnetic field, we quantitatively determine the relevant anisotropies, and demonstrate that the enhanced coercivity is related to the rotatable magnetic anisotropy of Fe$_{50}$Mn$_{50}$. We also demonstrate the consequences of the anisotropy changes with temperature.

\end{abstract}

\maketitle

Since its discovery almost 50 years ago~\cite{bean}, exchange-bias (EB) in ferromagnetic/antiferromagnetic (F/AF) bilayers has attracted much theoretical and experimental attention, but complete understanding remains elusive. Uncompensated fixed AF spins at the interface~\cite{bean}, AF domain walls either perpendicular~\cite{malozemoff} or parallel~\cite{mauri,mcmichael} to the F/AF interface, and spin-flop in AF at the interface with F~\cite{spinfloptheory} have been proposed as the dominant mechanisms, each claiming support by some experiments~\cite{bean,takano,stohrpinned,stohrspring,spinflop}. Furthermore, the AF magnetic order in proximity to F may be completely different from that of the bulk AF~\cite{calculations}, rendering some models oversimplified. The large number of models and the contradictory observations suggest a competition of the large number of energy scales involved in EB: anisotropies of F and AF, their exchange stiffness, thermal energies, and exchange coupling at the F/AF interface~\cite{nogues}.

In this work, we report new results addressing the origin of EB in an extensively studied Ne$_{80}$Fe$_{20}$(=Permalloy Py)/Fe$_{50}$Mn$_{50}$(=FeMn) bilayer. In contrast to most of the previous studies with polycrystalline samples, we used epitaxial (111) oriented films to avoid the effects of the AF granularity as discussed below. Unlike most studies of EB through hysteresis loop measurements with the magnetic field ${\mathbf H}$ along a fixed direction, we used a vector vibrating sample magnetometer in conjunction with a {\it rotating} magnetic field so that the angle $\alpha$ between ${\mathbf H}$ and the Py magnetization ${\mathbf M}$ can be measured. As we shall see, such measurements allow quantitative determination of anisotropy, and reveal features of EB not seen when the direction of ${\mathbf H}$ is fixed. Our data show that bilayers with thick FeMn exhibit only unidirectional anisotropy, small coercivity, and no hysteresis in response to a rotating ${\mathbf H}$. Whereas, in thin FeMn layers, the unidirectional anisotropy vanishes at room temperature (RT) $T=20^\circ$C, while the coercivity is enhanced, and more importantly a pronounced hysteresis in response to a rotating ${\mathbf H}$ appears. The unidirectional anisotropy and enhanced hysteresis generally coexist in EB systems~\cite{bean,nogues}. Our epitaxial samples enable us to separate and individually analyze these two distinct behaviors. By manipulating field and also temperature, we demonstrate that the coercivity and rotational hysteresis are associated with rotatable anisotropy, and explain these behaviors by a competition between the interfacial coupling and the bulk anisotropy of FeMn.

Epitaxial Cu(20)Py(8)/FeMn($t_{FeMn}$)/Cu(10) multilayers with $t_{FeMn}=2-50$ nanometers (nm) were made by sputtering on (0001) oriented Al$_2$O$_3$ substrates from alloy targets at RT in Ar pressure of 1.5 mTorr, in a vacuum chamber with a base pressure of $5\times 10^{-8}$ Torr. All thicknesses are in nm. The twinned epitaxial film growth was verified with x-ray diffractometry. The samples were exchange-biased by cooling  from $150^\circ$C to RT with $H=1$~kOe in the film plane. The measurements were performed at RT unless otherwise specified.

\begin{figure}
\includegraphics[scale=0.8]{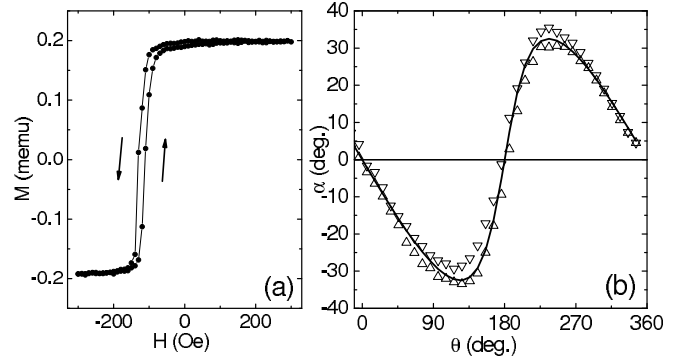}
\caption{\label{fig1} Magnetic response of the sample with FeMn(5). (a) Hysteresis loop, measured with positive ${\mathbf H}||{\mathbf H_B}$, with the arrows showing the scan direction. (b) The angle $\alpha$ between ${\mathbf H}$ and ${\mathbf M}$ {\it vs.} the angle $\theta$ between ${\mathbf H_B}$ and ${\mathbf H}$, with $H=200$~Oe. Upward (downward) triangles are for increasing (decreasing) $\theta$, and the solid curve is calculated using $H_B=112\pm10$~Oe, as described in the text.}
\end{figure}

Fig.~\ref{fig1}(a) shows the hysteresis loop of sample with FeMn(5), with ${\mathbf H}$ along the unidirectional anisotropy axis, the common way of measuring EB. The loop is characterized by an effective EB field ${\mathbf H_B}$, which gives the shift of the hysteresis loop, and the coercivity $H_c$, which is half of the difference of the fields at which the magnetization passes through zero, on the opposite reversal branches. Single Py films exhibit $H_c\approx0.5-1$~Oe. In contrast to the polycrystalline bilayers, where $H_B$ and $H_c$ are similar~\cite{parkin}, the EB effect of FeMn in Fig.~\ref{fig1}(a) is nearly "ideal" as simple models of EB predict~\cite{bean}, with a large $H_B$ and small $H_c$, also consistent with those of MBE-grown bilayers~\cite{epitaxial}.

The results for FeMn(5) sample using a rotating field $H=200$~Oe are shown in Fig.~\ref{fig1}(b), displaying the values of the angle $\alpha$ between ${\mathbf H}$ and ${\mathbf M}$ when the angle $\theta$ between ${\mathbf H_B}$ and ${\mathbf H}$ is varied. This variation of $\alpha$ is a direct  consequence of EB; Measurements for unbiased single Py films at similar $H$ gave $\alpha\approx0$. We note that the rotational anisotropy is unchanged when the chirality of the rotating field is reversed. During the rotation, the magnetic moment of Py, which we measured simultaneously with $\alpha$, deviated by no more than $10\%$ from its saturation value, characterizing the rotation of ${\mathbf M}$ as nearly monodomain. The solid line is a calculation based on the magnetic energy of~\cite{bean}
\begin{equation}\label{em}
E/M=-H\cos\alpha-H_B\cos(\alpha+\theta).
\end{equation}
which includes only unidirectional anisotropy $H_B=112\pm10$~Oe, and the Zeeman energy. Similarly to the hysteresis loop, the rotational measurement demonstrates a nearly "ideal" EB behavior, completely described by an effective field $H_B$. This property is independent of $H$ in the range of $30-1000$~Oe.

\begin{figure}
\includegraphics[scale=0.8]{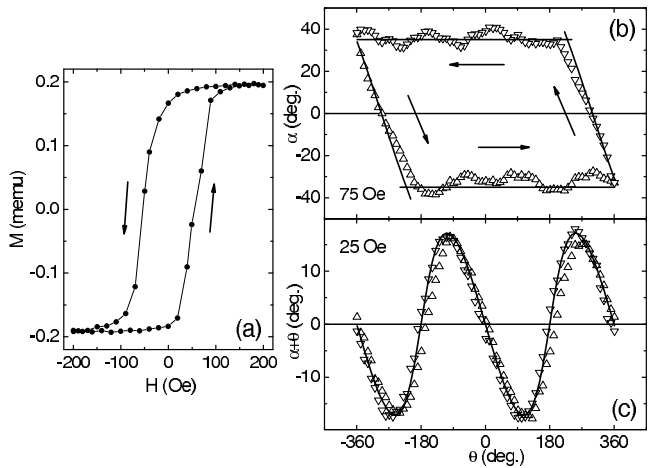}
\caption{\label{fig2} Data for the sample with FeMn(3). (a) Hysteresis loop with the arrows showing the scan direction. (b) The angle $\alpha$ {\it vs.} increasing (upward triangles) and decreasing $\theta$ (downward triangles), at $H=75$~Oe. The solid lines are calculated using $Q=43$~Oe for the horizontal lines, and $H^*_B=83$~Oe for the tilted lines, as described in the text. (c) $\alpha+\theta$ {\it vs.} increasing (upward triangles) and decreasing $\theta$ (downward triangles), with $H=25$~Oe. The solid curve is calculated using $H^*_B=83$~Oe, as described in the text.}
\end{figure}

In contrast to that of FeMn(5) sample, the hysteresis loop of sample with FeMn(3), shown in Fig.~\ref{fig2}(a), exhibits no EB but a large $H_c=50$~Oe. In rotating $H=75$~Oe, $\alpha$ first varies linearly, and then stays approximately constant during further field rotation, as shown in Fig.~\ref{fig2}(b). The direction of ${\mathbf M}$ always lags behind the rotating ${\mathbf H}$, giving rise to a large rotational hysteresis. This lag can be characterized by a torque exerted by FeMn on Py, opposite to the field rotation. The horizontal solid lines in Fig.~\ref{fig2}(b) are $\alpha=\pm\sin^{-1}(Q/H)$, with a constant $Q=43$~Oe. Thus, the effect of FeMn is equivalent to a torque field of $Q=43$~Oe applied perpendicular to ${\mathbf M}$. Only when the direction of the field rotation is reversed does the value of $\alpha$ change, as shown by the aperiodic cycle in 
Fig.~\ref{fig2}(b) for $-360^\circ<\theta<360^\circ$. However, when $H<50$~Oe, ${\mathbf M}$ is only reversibly disturbed, as shown in Fig.~\ref{fig2}(c) for $H=25$~Oe. In this case, the solid line is a calculation assuming unidirectional anisotropy as in Fig.~\ref{fig1}(b) with $H^*_B=83$~Oe. The anisotropy direction was determined from the best fit. The data at small $\alpha$ in Fig.~\ref{fig2}(b) can also be accounted for by assuming unidirectional anisotropy with the same $H^*_B=83$~Oe, as shown by sloped solid lines. These results show that FeMn(3) is magnetically ordered at RT and induces unidirectional anisotropy in Py. This anisotropy is broken by large reorientations of ${\mathbf M}$, giving no EB in the hysteresis loop.

\begin{figure}
\includegraphics[scale=0.9]{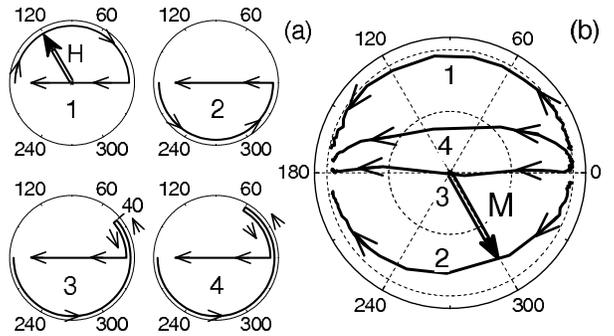}
\caption{\label{fig3} Results for FeMn(3) sample. (a) Polar coordinate schematics of the four combinations of field rotations and linear scans, as described in the text. The angular coordinate is $\theta$, the radial coordinate is $|{\mathbf H}|$, not to scale. Arrows show the direction of the field variation. A double arrow in panel 1 shows the rotating ${\mathbf H}$ at $\theta=120^\circ$. (b)  Polar plots of the magnetization reversal initiated by the field scans shown in (a). The radial coordinate is $|{\mathbf M}|$, and the angular coordinate is $\alpha$.
A double arrow shows ${\mathbf M}$ in case 2, oriented at $\alpha=300^\circ$ during the reversal.}
\end{figure}

To demonstrate that the results of Fig.~\ref{fig2} are due to the rotatable anisotropy of FeMn, we performed four different combinations of field rotations, followed by linear scans, schematically illustrated in Fig.~\ref{fig3}(a). In the first case, the field was rotated clockwise by $180^\circ$, in the second it was rotated counterclockwise, in the third and the fourth cases it was also first rotated counterclockwise, and then clockwise by $40^\circ$ and $60^\circ$, correspondingly. In all four cases, the rotating field of $1$~kOe was much larger than the saturation field of $\approx 150$~Oe, ensuring that ${\mathbf M}$ closely followed ${\mathbf H}$. After the rotations, the magnetization was reversed by scanning $H$ from $200$~Oe to $-200$~Oe along a fixed axis, the same in all four cases. The resulting magnetization reversal trajectories are shown in Fig.~\ref{fig3}(b) in polar coordinates. The four different initial rotations lead to strikingly different reversal modes during the same subsequent linear field scans. In the first case, the reversal occurred through a nearly monodomain counterclockwise rotation, opposite to the initial field rotation. In contrast, in the second case, magnetization reversed through a clockwise rotation. In case 3, the reversal occurred through an inhomogeneous state with average $M=0$ (the center of the polar coordinates), and no overall rotation, giving an approximately horizontal line in Fig.~\ref{fig3}(b). In the fourth case, the magnetization reversed through an overall counterclockwise rotation. The results of Fig.~\ref{fig3}(b) demonstrate that the magnetization reversal by linear field scanning, the common technique of characterizing EB, does not reveal the intrinsic properties of EB; In Fig.~\ref{fig3}(b) the magnetization reversed through different modes determined by the sample history. On the other hand, the results of Fig.~\ref{fig2} for rotating field were independent of the sample history, revealing the intrinsic features of anisotropy.

The data of Figs.~\ref{fig2},~\ref{fig3} indicate that the unidirectional anisotropy axis of FeMn(3) starts to rotate when its angle $\psi$ with ${\mathbf M}$ exceeds a critical value $\psi_c$. For a fixed unidirectional anisotropy, we defined $\theta$ as the angle between the anisotropy direction and ${\mathbf H}$, so that $\psi=\alpha+\theta$. For small $H$ as in Fig.~\ref{fig2}(c), $\alpha+\theta<\psi_c$, the anisotropy axis is not reoriented, giving only reversible behaviors. However, when $\alpha+\theta>\psi_c$ at $\theta\approx \pm-290^\circ$ for decreasing/increasing $\theta$ in Fig.~\ref{fig2}(b), the anisotropy axis is subsequently dragged by the rotating $M$ at $\psi_c\approx \Delta\theta-\alpha=35^\circ$, where $\Delta\theta$ is the field rotation angle from the collinear orientation with $M$ to the onset of the drag. On the other hand, from Figs.~\ref{fig2}(b,c) $\sin^{-1}(Q/H^*_B)=31^\circ$ is close to $\psi_c$. Thus, the rotational torque, exhibited in Fig.~\ref{fig2}(b), is induced by the same FeMn spins that give the unidirectional anisotropy in Fig.~\ref{fig2}(c), but now following the rotating ${\mathbf M}$. After the initial field rotation in the first two cases in Fig.~\ref{fig3}, ${\mathbf M}$ rotates towards the anisotropy axis, opposite to the initial rotation. In the third case shown in Fig.~\ref{fig3}, the counterclockwise rotation to $\theta\approx 40^\circ$, close to $\psi_c$, brings the anisotropy direction to $\theta\approx 0$, giving no overall rotation of ${\mathbf M}$ during the subsequent reversal. Finally, in the fourth case in Fig.~\ref{fig3}, the initial rotation brings the anisotropy axis to $\theta>0$, leading to reversal through an overall rotation counterclockwise.

Samples with $t_{FeMn}\ge5$~nm showed only unidirectional anisotropy, and small hysteresis in rotating field measurements, similar to that shown in Fig.~\ref{fig1}. They did not exhibit a history dependence of the magnetization reversal by linear field scans, such as shown in Fig.~\ref{fig3}(b). FeMn(2) sample did not exhibit EB or rotational hysteresis at RT. In contrast to epitaxial samples with $t_{FeMn}\ge5$~nm, polycrystalline Py(8)/FeMn($t_{FeMn}$) samples with $t_{FeMn}\ge5$~nm showed both unidirectional anisotropy and significant rotational hysteresis, correlated with enhanced $H_c$. Thus, our epitaxial samples let us separately study the features of anisotropy, generally coexisting, and obscuring each other, in polycrystalline samples.

The results shown in Figs.~\ref{fig1},~\ref{fig2} can be explained by comparing the important energy scales. To determine whether a partial domain wall forms in FeMn when ${\mathbf M}$ is rotated, we compare the AF domain wall energy to the strength of EB~\cite{neel,mauri,mcmichael}. If a domain wall were present in FeMn(5), its width would necessarily be less than $5$~nm to avoid unwinding from the side of AF opposite to Py, which would erase EB and give hysteretic behaviors~\cite{neel,mcmichael}. Using $5$~nm for its width, and $A\approx 10^{-11}$~J/m for FeMn exchange stiffness, the AF domain wall energy per unit area is $E_{dw}\approx 2\times10^{-2}$J/m$^2$ for the $180^\circ$ wall, about two orders of magnitude larger than the unidirectional anisotropy energy $E_B=2Mt_{Py}H_B=1.5\times10^{-4}$~J/m$^2$ for $H_B=112$~Oe. Thus, the observed EB is due to the uncompensated fixed FeMn spins at the interface. Using $10^{-21}$J for exchange energy per interfacial atom, we estimate the uncompensated spin density of about $\approx0.4\%$, consistent with the x-ray dichroism studies of the interface~\cite{malozemoff,stohrpinned}. Since $E_{dw}\gg E_B$, the FeMn spins do not form a domain wall, they either remain fixed or rotate nearly coherently throughout the FeMn thickness. Thus, the dependencies of the samples' behaviors on $t_{FeMn}$ must be determined by the competition between the Py/FeMn exchange coupling energy (independent of $t_{FeMn}$) and the volume FeMn anisotropy (proportional to $t_{FeMn}$). In samples with $t_{FeMn}\ge5$, the FeMn anisotropy dominates. The FeMn spins are then pinned, giving Heisenberg exchange energy consistent with Eq.~\ref{em}. At $t_{FeMn}<5$, the FeMn anisotropy becomes smaller than the interfacial exchange, so rotating ${\mathbf M}$ also rotates the FeMn spins, giving irreversible behaviors. The FeMn spins start rotating as the torque exerted by ${\mathbf M}$ exceeds a critical value $Q_c=H^*_B\sin(\psi_c)$, below which they remain fixed. Hysteresis in polycrystalline samples with large $t_{FeMn}$ suggests small FeMn crystallines at the interface, giving anisotropy features similar to epitaxial samples with small $t_{FeMn}$~\cite{parkin}.

\begin{figure}
\includegraphics[scale=0.8]{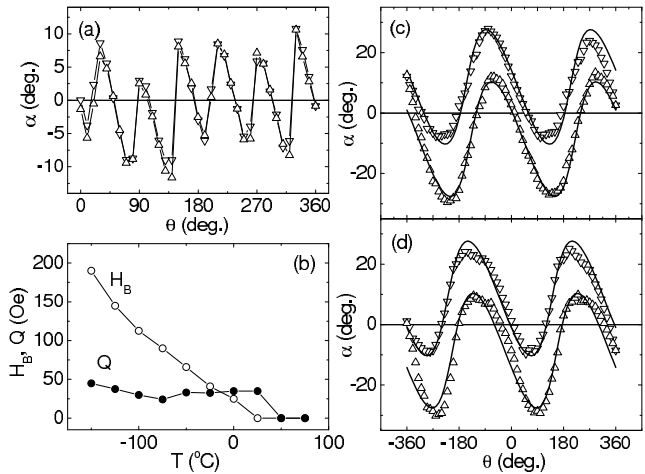} 
\caption{\label{fig4} Data for FeMn(3) sample. (a) $\alpha$ {\it vs.} $\theta$ at $75^\circ$C and $H=50$~Oe. (b) $H_B$ (solid symbols) and $Q$ (open symbols) {\it vs.} $T$. (c) $\alpha$ {\it vs.} $\theta$ at $-50^\circ$C and $H=200$~Oe, after $180^\circ$ to $0^\circ$ rotation of $H=1$~kOe at $25^\circ$C, with subsequent cooling from $25^\circ$C at $H=1$~kOe. The solid curves are calculations using $H_B=66$~Oe, $Q=30$~Oe, and easy axis at $\theta=32^\circ$. (d) same as (c), but the rotation at $25^\circ$C is from $-180^\circ$ to $0^\circ$ $H=1$~kOe, and the calculations are with the easy axis at $\theta=-32^\circ$. In (a,c,d), upward (downward) triangles are for increasing (decreasing) $\theta$.}
\end{figure}

The rotatable anisotropy of FeMn(3), demonstrated in Figs.~\ref{fig2},~\ref{fig3}, also enables us to analyze the mechanisms of the temperature dependence of EB. As shown in Fig.~\ref{fig4}(a), the rotational hysteresis of sample N3 is replaced at $T=75^\circ$C by a reversible 6-fold anisotropy. This anisotropy is induced by the magnetically ordered FeMn(111), as it disappears at $T\approx125^\circ$C, which is close to the bulk N\'{e}el temperature of FeMn. These data are inconsistent with the model, explaining the decrease of the blocking temperature in thin AF layers by reduction of their ordering temperature~\cite{finitesize}. As Fig.~\ref{fig4}(b) shows, EB appears in FeMn(3) sample at $T<20^\circ$C. $H_B$ linearly increases as $T$ is reduced, while $Q$ only weakly depends on $T$, with a weak maximum at the onset of EB. The approximately linear dependence of $H_B$ on temperature is quite generally seen in EB systems~\cite{parkin}. It has been alternatively attributed to the variation of AF anisotropy~\cite{parkin,fluctuation,mcmichael}, or the changes of uncompensated AF spin density at the interface~\cite{takano}. 

We used two different field-cooling procedures to demonstrate that the FeMn anisotropy must increase at lower temperatures, leading to "freezing" of the FeMn(3) spins rotatable at RT. First, we rotated ${\mathbf M}$ at RT with $H=1$~kOe, clockwise in the first case, and counterclockwise in the second. We subsequently cooled the sample to $-50^\circ$C with $H=1$~kOe oriented at $\theta=0$ in both cases, and measured the anisotropy using a rotating field. In the first case, shown in Fig.~\ref{fig4}(c), the anisotropy axis was oriented at $\theta\approx 32^\circ$. In the second case, shown in Fig.~\ref{fig4}(d), the anisotropy axis was at $\theta\approx-32^\circ$. These values of $\theta$ are close to our estimate for $\psi_c$, with which the FeMn spins lag behind the rotating ${\mathbf M}$ at RT. Thus, the EB direction is not set by the direction of ${\mathbf M}$ or the field direction during the cooling, but rather by the FeMn spin anisotropy before the cooling. These data are inconsistent with the models assuming that the increase of $H_B$ at low-temperature is associated with the blocking of spins thermally fluctuating at higher $T$~\cite{parkin,fluctuation}. Such models would predict the low-$T$ anisotropy to be in the same direction as ${\mathbf M}$ during the field-cooling. Instead, our data show that spins with a certain rotatable orientation become fixed at lower temperature due to the increase of FeMn anisotropy, and thus contribute to EB.

In FeMn(2) sample, the rotational hysteresis similar to Fig.~\ref{fig2}(b) appeared below $T\approx -50^\circ$C, and EB appeared at lower $T$. In FeMn(5) sample, hysteresis appeared at $T\approx 70^\circ$C, near its blocking temperature.  However, samples with $t_{FeMn}>5$~nm did not show an enhanced hysteresis near their blocking temperatures. These results, as well as data for polycrystalline samples, indicate that rotatable anisotropy is a general feature of thin FeMn layers and small crystalline grains. 

In summary, EB in (111)-oriented epitaxial Py/FeMn bilayers is due to the interaction between the  fixed uncompensated interfacial FeMn spins and the Py magnetization. The onset of spin rotations in thin FeMn layers is associated with rotational hysteresis and enhancement of coercivity, due to the  rotatable anisotropy of FeMn. With variable temperature measurements, we demonstrate that rotatable anisotropy becomes "frozen" at low temperature, contributing to EB.

We acknowledge helpful discussions with M.D. Stiles and V.I. Nikitenko, and support from the NSF through Grant DMR00-8031.

\end{document}